# Electrical Stressing Induced Monolayer Vacancy Island Growth on TiSe$_2$

Husong Zheng[1#], Salvador Valtierra[2#], Nana Ofori-Opoku[4,5#], Chuanhui Chen[1], Lifei Sun[3], Liying Jiao[3], Kirk H. Bevan[2*], and Chenggang Tao[1*]

[1]Department of Physics, Virginia Tech, Blacksburg, Virginia 24061, USA

[2]Materials Engineering, McGill University, Montreal, Quebec H3A 0C5, Canada

[3]Department of Chemistry, Tsinghua University, Beijing 100084, China

[4]Materials Measurement Laboratory, National Institute of Standards and Technology, Gaithersburg, MD 20899, USA

[5]Center for Hierarchical Materials Design, Northwestern University, Evanston, IL 60208, USA

# These authors contributed equally to this work.

* Corresponding to: kirk.bevan@mcgill.ca and cgtao@vt.edu.

ABSTRACT

To ensure the practical application of atomically thin transition metal dichalcogenides, it is essential to characterize their structural stability under external stimuli such as electric fields and currents. Using vacancy monolayer islands on TiSe$_2$ surfaces as a model system, for the first time we have observed a shape evolution and growth from triangular to hexagonal driven by scanning tunneling microscopy (STM) electrical stressing. The size of islands shows linear growth with a rate of $(3.00 \pm 0.05) \times 10^{-3}$ nm/s, when the STM scanning parameters are held fixed at $V_s = 1.0$ V and $I = 1.8$ nA. We further quantified how the growth rate is related to the magnitude of the tunneling current. Our simulations of monolayer island evolution using phase-field modeling are in good agreement with our experimental observations, and point towards preferential edge atom dissociation under STM scanning driving the observed growth. The results could be potentially important for device applications of ultrathin transition metal dichalcogenides and related 2D materials subject to electrical stressing under device operating conditions.

## I. INTRODUCTION

Emerging 2D materials, such as atomically thin transition metal dichalcogenides (TMDs), have been the subject of intense research due to their fascinating properties and potential practical applications [1-3]. TMDs have a layered structure in which a transition metal atom layer is sandwiched between two chalcogen atom layers and adjacent layers are stacked via Van der Waals forces. Atomically thin TMDs vary from metallic, semimetallic, to semiconducting, and can be used in electronic devices, phototransistors, solar cells and gas sensors [4-9]. As a member of the TMDs, 1$T$-TiSe$_2$ is a widely studied charge density wave (CDW) material. Below 200 K, 1$T$-TiSe$_2$ undergoes a phase transition to a CDW state, showing a 2 × 2 × 2 commensurate superlattice [10-12]. Such a transition in 1$T$-TiSe$_2$ implies great potential applications in optoelectronics [13], voltage-controlled oscillators [14] and ultrafast electronics [15].

However, the properties of atomically thin TMDs and other 2D materials are both sensitive to and governed by defects and interfaces, such as domain boundaries and edges [16-19]. Usually these interfaces are more susceptible to thermal fluctuations and external stimuli, than the bulk of the material [20,21]. Thus, for practical applications of mono- and few-layer TMDs, it is essential to characterize their structural stability when subjected to the external stimuli present in devices, such as: electrical fields, irradiation and other forcing conditions. In situ investigations of said systems are usually difficult given the dynamical conditions, and so far quantitative characterizations of structural stability of these systems are still lacking. Furthermore, it is important to utilize theoretical modeling to interpret the results and understand the impact of measurement parameters on the observed trends. Numerical techniques such as molecular dynamics (MD) and most recently phase-field modeling enable one to calculate the dynamical evolution of materials. However, the phase-field method, unlike MD, can explore the diffusive time and length scales appropriate for studying microstructure evolution in electronic materials.

Using monolayer vacancy islands on titanium diselenide (TiSe$_2$) surfaces as a model system as shown in Fig. 1, we experimentally and theoretically investigated their shape evolution and growth rate driven by scanning tunneling microscopy (STM) electrical stressing. The equilibrium triangular monolayer vacancy islands evolve to a hexagonal shape and the island area shows a non-linear area growth dependence with respect to time (when electrically stressed by a STM tip). The growth rate dependence on the tunneling current is experimentally determined. Our simulations of shape and size evolution using a phase-field model are consistent with our experimental observations, and suggest that the STM driven vacancy island growth is driven by island edge atom dissociation under electrical stressing. The results from this study could be potentially important for understanding the device applications and limitations of atomically thin TMDs and other 2D materials subject to electrical stressing. Particularly, for TMD transistors subject to analogous vertical electric fields and leakage tunneling currents through gating.

## II. METHODS

### A. Experimental methods

The single crystals of 1$T$-TiSe$_2$ were grown with chemical vapor transport (CVT) using iodine as the transport agent purchased from 2D age. Before scanning tunneling microscopy (STM) characterization, our X-ray diffraction (XRD) and electrical measurements confirmed the high quality of the crystals.

The STM experiments were carried out in a customized STM (Omicron STM1). The fresh surfaces of samples were prepared by the mechanical exfoliation method. STM scanning was performed at room temperature in a vacuum chamber with a base pressure of lower 10$^{-9}$ Torr. The STM tips used in the experiments were etched from polycrystalline tungsten wires. All the STM images were obtained in the constant current mode.

### B. Theoretical simulation method

Phase-field modeling is a phenomenological technique capable of simulating the kinetics of phase transformations and microstructural evolution. This technique relies on the construction of a free energy functional to describe the thermodynamics of the system of interest. In our phase-field model, a binary component system was used with two components (atoms and vacancies) [22,23]. The current simulations are based on that developed by Rokkam, El-Azab, Millet and Wolf [24] in which an order parameter ($\phi$) and a concentration of vacancies ($c_v$) describe the system free energy (F) in the following form:

$$\mathcal{F}[\phi, c_v] = N \int [\, G(c_v)h(\phi) + w(c_v, \phi) + \tfrac{1}{2}|\varepsilon_\phi \nabla \phi|^2 + \tfrac{1}{2}|\varepsilon_v \nabla c_v|^2 ]\, dV \quad (1)$$

The variable $G(c_v)$ represents the enthalpic and entropic contributions, whereas $h(\phi)$, $p(\phi)$ and $g(\phi)$ represent interpolation functions that allow for smooth transitions of bulk free energy information across interfaces. They drive the free energy terms to zero as $\phi$ goes to $0$ or $1$. The free energy admits two stable bulk solutions by design. At $\phi = 0$, we have a stable well that represents the equilibrium concentration of vacancies determine by $G(c_v)$; and at $\phi = 1$, the stable well is that for a void at $c_v = 1$ given by $w(c_v, \phi)$, as shown in Fig. 2. Finally, $\varepsilon_v$ and $\varepsilon_\phi$ represent the gradient energy coefficients for the order parameter and concentration fields respectively, which determine the properties of interfaces. Further information about the thermodynamics has been provided in the third section of the supplementary material. The governing equations driving the system evolution are:

$$\frac{\partial \phi}{\partial t} = -M_\phi \frac{\delta \mathcal{F}}{\delta \phi} \quad (2)$$

$$\frac{\partial c_v}{\partial t} = -\nabla \cdot J + Zg(\phi) \quad (3)$$

$$J = -M_v \nabla \frac{\delta \mathcal{F}}{\delta c_v} \quad (4)$$

$$M_v = \frac{D_v\, c_v^2 (1 - c_v)^2}{K_B T} \quad (5)$$

$$g(\phi) = (\phi - 1)^2\, e^{-18.42(\phi - 1)^2} \quad . \quad (6)$$

In the above equations, $M_\phi$ represents the interfacial mobility of the system and $M_v$ is the diffusional mobility which is related to the diffusion coefficient ($D_v$) of vacancies. This mobility can be concentration and order parameter dependent; in this work we only allow it to have a compositional dependence. The term $Zg(\phi)$ represents a source term driving the preferential formation of vacancies at void edges, via a perturbing current flow from a STM tip.

The kinetics of the model follow from dissipative thermodynamics of conserved and non-conserved fields. Namely, the phase-field in Eq. (2) follows the dynamics of free energy minimization, so-called Allen-Cahn dynamics; while, the concentration field (setting the source term to zero) follows the continuity equation of flux conservation Eq. (3), so-called Cahn-Hilliard dynamics. These dynamics are also called *Model A* and *Model B* respectively, and when combined are referred to as *Model C*.

The original model used was set in dimensionless units, by reverting some of the scaling, the model is then used to simulate void growth [24]. To capture the observed experimental hexagonal anisotropy, the order parameter gradient coefficient is made to be orientation dependent in the form:

$$\int |\varepsilon_\phi \nabla \phi|^2 \, dV \rightarrow \int |\breve{\varepsilon}_\phi(\theta) \nabla \phi|^2 \, dV \qquad (7)$$

Where $\theta$ is the orientation of the interface normal referenced to the lab frame, $A(\theta)$ modulates the anisotropy of the order parameter gradient coefficient, and $\varepsilon$ describes the degree of anisotropy of the surface tension. Further explanation regarding the generation of the anisotropy in this model has been provided in the third section of the Supplementary Material along with the parameters used for the simulation.

## III. RESULTS AND DISCUSSIONS

The atomic structure of 1$T$-TiSe$_2$ is schematically shown in Fig. 1a. Fig. 1b displays a large scale STM image of a freshly cleaved TiSe$_2$ surface that appears atomically smooth. From the atomically resolved STM images (insets in Figs. 1b and 1c), we were able to determine the lattice constant is 3.54 ± 0.06 Å. Some point defects, such as Se vacancy defects, as shown in the inset of Fig. 1b, were also observed. The intrinsic atomic defects have been previously observed in similar systems [12,25]. After annealing at 350℃ for 2 hours in the UHV chamber with a base pressure of lower 10$^{-9}$ Torr, triangular vacancy islands with a size ranging from 10 to 30 nm were formed in a random distribution on the surface as shown in Fig. 1c. The density of islands was determined to be (5.0 ± 2.6) × 10$^{10}$ cm$^{-2}$. The depth of islands, as shown in the line profile in Fig. 1d, was measured at 6.2 ± 0.3 Å, consistent with the monolayer height of 1$T$-TiSe$_2$ [26]. By comparing with atomically resolved STM images obtained from small scan areas (see the inset of Fig. 1c), the island edges can be seen to align overall along the highly symmetric orientations of the surface with some small deviation at short segments. The triangular shape observed is that commonly displayed for monolayer vacancy islands on TMD surfaces. Similar monolayer vacancy islands have previously been observed in other TMDs after annealing [27,28].

When an STM tip is continuously scanning, and thereby electrically stressing the surface, our results show that the monolayer vacancy islands grow and their shape changes. Figs. 3a-h provide a snapshot of images spaced at a time interval of ~ 800 minutes, selected from a total data set of 974 sequential images, which clearly display the shape of the monolayer vacancy island evolving from triangular to hexagonal under STM electrical stressing. Such growing behavior only happens when an island under continuous scanning (see Figs. S1 and S2 in the Supplementary Material). To aid visual comparison, the images in Figs. 3a-h were cropped from the scanning area of 88 × 88 nm$^2$. The scanning rate is 500 s per image with a 256 × 256 pixel resolution and a line-scanning speed of 90.9 nm/s. To extract the area of the monolayer vacancy islands, the perimeters were identified as the point at which the surface height is 30% of the step height lower than the top layer. This threshold was chosen to minimize the error included by island edge roughness, pixel noise, and STM image noise. The accuracy of the island area is given by the step width and is typically less than 3 nm for a sharp STM tip. To validate the analysis, we carefully checked the obtained island perimeters by visual inspection and ensured that the island area deviations introduced by the STM image processing computer code was at most 1%. The vacancy island area uncertainty was determined by varying the threshold values from 10% of the step height lower than the top layer to the midway between the top layer and the bottom of the monolayer vacancy islands. Fig. 4 is a plot of the square root of the area of the monolayer vacancy island shown in Figs. 3a-h as a function of time. The plot clearly indicates a non-linear parabolic dependence of the island area with respect to the STM scan time. By fitting the plot in Fig. 4, we obtained a constant growth rate of $(3.00 \pm 0.05) \times 10^{-3}$ nm/s.

To interpret our experimental observations, we employed a phase-field model to simulate the island growth and shape evolution. At first, we attempted to model the vacancy island growth by assuming a uniform bulk distribution of vacancies induced by STM electrical stressing of the substrate. This yielded out-of-equilibrium vacancy-void clustering results that followed the expected linear area growth with respect to time, in line with the diffusional growth law for bulk vacancy clustering [29]. The experiments, however, demonstrated a non-linear behavior (as shown in Fig. 4). We therefore concluded that electrical stressing applied by the STM tip must be preferentially driving vacancy formation at the vacancy island edges, due to the fewer number of bonding neighbors present for edge atoms, as discussed below.

The behavior of the STM tip on the material was examined through phase-field modeling in order to simulate the phenomena that generated vacancies in the dichalcogenide [30-32]. Our phase-field model evolves according to the concentration of vacancies in the system, where a diffusion equation [see Eq. (3)] tracks the evolution of this field. The effect of electrical stressing from the STM tip was simulated in the model as a source term that is included in the concentration field equation of motion given by Eq. (3). This source term is added to generate a high concentration of vacancies at interfaces and a very small amount of vacancies in the bulk of the material. Due to the diffusivity of the vacancies being so low and the contribution of a surface diffusion term $c_v^2(1-c_v)^2$ in the mobility, vacancies in the bulk are for practical purposes immobile while vacancies at the interface or in contact with the interface have a higher mobility thus allowing the void to take vacancies and grow. This low bulk diffusion has been validated experimentally by very low measured diffusion constants in dichalcogenides [30]. From said work, we infer an estimate of the diffusion coefficient for vacancies in the TiSe$_2$ bulk. It is also discussed how vacancy mobility is influenced by the presence of vacancy clusters. This is relevant to our model, since this determines the concentration dependence of the mobility term in equation 7. The vacancy mobility involves the migration barrier for vacancies surrounded by atoms of the material and near a vacancy cluster. This leads to the argument that vacancy clusters provide a thermodynamic driving force that makes it easier for vacancies to move near clusters.

This source term which is meant to imitate electron injection, at a coarse-grained level, in a material to generate vacancies (due electrical stressing of the material by the STM tip) was set to a constant Z multiplied by a function g($\phi$) that takes the form of a decaying exponential starting from the interface [see Eq. (3)]. The scope of the model is to capture the behavior of vacancies in a material. This behavior is limited to bulk diffusion, coarsening and surface diffusion of vacancies. The source of excess vacancies in the system can be stated as either an initial excess of vacancies or through the source term Zg($\phi$) in equation 5. As for the source term, it is phenomenological in essence. However, it was constructed to mimic the effect from the STM tip. One limitation of the model is that if some thermodynamic process had the same behavior as the one constructed, the model would not be able to distinguish between the vacancies from the tip and the other thermodynamic process. The Z term controls the amount of vacancies being added to the material while g($\phi$) controls the ratio of vacancies between the interface and the bulk. That is, it controls the fact that vacancies are preferentially generated at the edge of a void while the bulk is less likely to form vacancies. In this manner the source term generates a higher amount of vacancies at interfaces and a lower amount of vacancies in the bulk of the material, thereby yielding a non-linear growth when the area is plotted against the simulation time as seen through the good agreement between theory and experiment in Figs. 3 and 4. Again, the vacancy island edge source term is physically justified by the fewer number of bonding neighbors present for interface adatoms and therefore a lower kinetic barrier for interface adatom dissociation (as compared to the bulk) upon electrical stressing by the STM tip. And it is only through preferential edge vacancy generation that we are able to explain the non-linear vacancy island area growth with respect to time.

Although similar vacancy growth has already been seen on some TMDs through both STM and AFM measurements [33-42]. It is the first time that a direct correlation between preferential electrical stressing at vacancy island edges and the observed growth rate has been established. Moreover, the observed hexagonal shape transition under non-equilibrium stressing is also unique. We believe that the triangular vacancy islands we observe from the annealing leave the tip and edge atoms highly vulnerable and ejectable from their position. It is also known that while triangular islands have one type of zigzag edge, hexagonal islands possess two types of zigzag edges. The chalcogen ejection could lead to the formation of both types of edges. This structure itself becomes stable due to the removal of chalcogen atoms from the edges. This then modifies the edge chemistry leading to a change of surface energetics. More specifically, the surface energetics we refer to are the free energy contributions of the vertices and of the boundary which depend on the chalcogen chemical potential. Thus, the shape evolution can be understood from the selenium vacancy generation at the edge perspective. Another factor that can influence the shape is the amount of chalcogen. Lauritsen [18] and Wang [43] used different methods to prove that MoS2 will develop a hexagonal shape under sulfo-reducing conditions while sulfur-rich conditions lead to triangular islands. Thus, we can infer that the available amount of chalcogen (selenium in this case) plays a key role in the island growth. The hexagonal growth observed in the experimental images was simulated, as shown in Fig. 3i-p, by modifying the gradient energy coefficient of the order parameter in Eq. (1) [22]. This was achieved by generating an angular dependence which makes the void clusters grow in preferential directions. Again, by slightly modifying the angular dependence in Eq. (1) it is possible to shift the orientation of growth. This was done to capture the crystalline anisotropy present in $TiSe_2$ which doesn't automatically arise from the chosen free energy functional in Eq. (1).

Now we can apply our model to capture the isolated growth of single voids and correlate the model directly with the observed STM image results in Fig. 3a-h. As an initial condition, we start with a void (i.e., a cluster of vacancies) in the initial shape of a triangle like those of the experiment after annealing as shown in Fig. 3i. We also assume that the interaction between voids is negligible and set the bulk vacancy concentration close to the equilibrium value for the given conditions. By allowing the phase-field model to progress at diffusional time scales we obtained the results in Fig. 3i-p, which is in good agreement with the experimental results in 3a-h both in terms of area and the structural transition from triangular to hexagonal.

In the simulation, the observed experimental non-linear growth in the vacancy island area (shown in Fig. 4) is captured by a preferential interfacial source term, as discussed above and in the context of Eq. (3). Physically, this can be understood as inducing interfacial diffusion that occurs at a faster rate than bulk-diffusion limited growth which would produce a linear trend when the area is plotted against the time, rather than the non-linear parabolic growth trend shown in Fig. 4. Only by inducing vacancy formation preferentially at the void edge can we capture the parabolic growth trend with respect to time in our model.

As long as there is some kind of preferential adatom dissociation at the interface, the model produces a parabolic growth behavior. Several functions were used, that produced very similar results, however a more detailed analysis of how changing the function affects the non-linear growth behavior is left for future work. Importantly, the very small diffusion coefficient assumed for TiSe$_2$ only allows the closest vacancies to the interface to "travel" towards the void and contribute to the growth leading to interfacial diffusion as the key growth mechanism. Physically speaking, this is meant to represent an etching process in which atoms are removed from the edge of the void due to electrical stressing from the STM tip. However, the model not only includes vacancy generation at the edge of a void (which is meant to represent the etching) it also draws any vacancies that are very close to the void. This is justified through thermodynamic arguments in which excess vacancies want to cluster in order to minimize the free energy (which is also a way of reducing strain on the lattice). Also that the migration barrier in dichalcogenides that normally makes diffusion much slower in comparison to metals, becomes quite low near a cluster of vacancies. All vacancies that are near the void then, experience a thermodynamic potential drawing them towards the closest void.

To understand the role of the tunneling current in the growth process, and further establish that the vacancy island growth is driven by electrical stressing by the STM tip, we measured monolayer island growth at various tunneling currents. Fig. 5a shows a variation in island growth rates as the tunneling current is varied from 0.3 nA to 1.8 nA. In the measurements, the scanning areas, rates and sample bias were kept the same for the entire data set. The scanning area was held at 58.52 × 58.52 nm$^2$ and the scanning rate at 60.63 nm/s, with the sample bias set at 1.0 V. We were able to obtain an excellent fit to the parabolic area growth at different tunneling currents by altering only the source constant Z magnitude in the phase-field model as shown in Fig. 5a. Only the initial island size was taken from experimental data in the phase-field results presented in Fig. 5a. From this fit we were also able to obtain the growth rate as a function of the tunneling current, as shown in Fig. 5b. There is an approximate quadratic non-linear relationship between growth rate and the tunneling current.

At this point we cannot conclusively state whether the quadratic non-linear relationship between the growth rate and the tunneling current is driven by the electric field or electric current increase (or both). However, in our scanning current variation the bias is held constant at 1 V, this implies that the current gain arises from modulation of the tunneling barrier width ($W$). If one assumes a simple tunneling barrier where the current may be approximately described by the proportionality relation $I_{tip} \propto \exp(-2\kappa W)$, where $\kappa = \sqrt{2mU}/\hbar$ – $m$ is the mass of an electron, $U$ is its tunneling barrier, and $\hbar$ is Planck's constant. Then the variation in current from 0.3 nA to 1.8 nA can be attributed to an approximate 1 Å variation in the tunneling barrier width, assuming $U \approx 4$ eV. Under such small distance perturbations, the increase in the electric field between the tip and sample is not likely to be a major factor (given that applied bias is held constant in our measurements). From this we conclude that the rate of vacancy generation is most likely directly correlated to the amount of electrons injected per unit time. Since the tip bias is held constant, the mean energy per tunneling electron remains the same, however the number of tunneling electrons increases as the barrier width (resistance) is reduced. It seems logical to conclude that the corresponding increase in the power dissipated by the tunneling electrons ($P \propto I_{tip}^2$) leads to a corresponding proportional increase in the rate of energy dissipated in the sample through bond breaking at the void edges and the corresponding parabolic increase in the etching rate with respect to current flow (as shown in Fig. 5b).

However, the mathematical relationship between the vacancy generation source term magnitude (Z) and the phase-field simulated growth rate does not display a parabolic behavior, as the growth rate vs current plot from the experimental data does (see Fig. 5b). A mathematical operation was then performed to compare against the parabolic behavior of the growth rate vs current. This operation, shown in Fig. 5b, evaluates the source value (Z) to the power ¼. At this point it is difficult to establish a relationship between the vacancy source magnitude and the tip current without more exhaustive first-principles kinetic and electron transport simulations [44,45]. The reason behind this difficulty is that since the distance between the sample and the STM tip does not remain constant, the interactions between the tip and the surface of the sample are quite difficult to establish precisely. A detailed first-principles analysis of this rich and intriguing physics is left for future work.

## IV. CONCLUSIONS

In this work, we observed a non-linear area growth (or linear size growth) behavior with respect to time and a shape evolution from triangular to hexagonal in monolayer vacancy islands on $TiSe_2$ surfaces under STM scanning. The growth rates were measured at various tunneling currents and determined to be directly correlated to electrical stressing by the STM tip. A phase-field model is used to simulate the growth behavior of a triangular void that is formed during the annealing of the $TiSe_2$ and its diffusional growth when exposed to electrical stressing by a STM tip. There is good agreement between the observed experimental results and the predictions from our simulation by using a vacancy source term preferentially acting at the void edges in the equation of motion of the vacancy concentration field. However, a deeper understanding of the governing kinetics and electron transport phenomena requires further detailed first-principles simulations, since the coarse-grained continuum field applied in our analysis does not provide the required atomistic insight about the formation of interfacial vacancies when a material is exposed to electrical stressing by an STM tip. Nevertheless, the model demonstrates that the observed growth rates are driven by interfacial adatom dissociation through excitation by an STM tip. These results could likely impact our understanding of the reliability of TMDs in electronic devices, when exposed to electrical stressing in realistic operating conditions, including vertical gating fields and tunneling currents in novel TMDs transistor designs (analogous to the vertical STM induced stressing studied in this work).

**Figures**

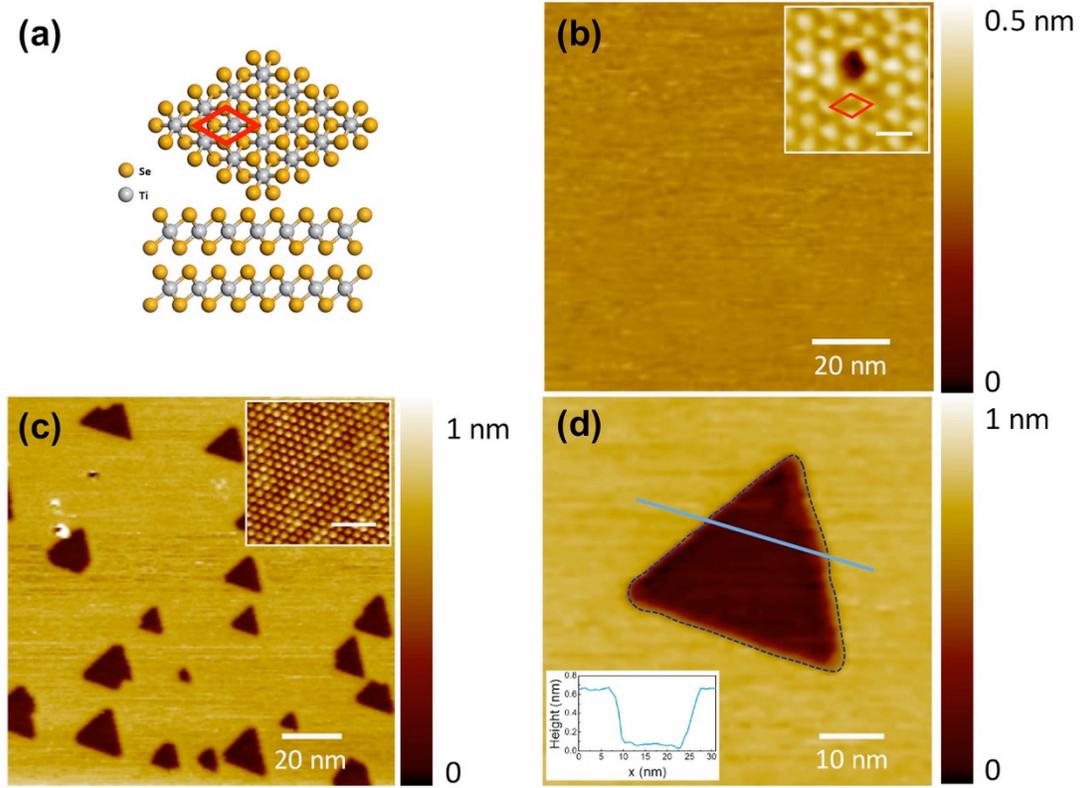

FIG. 1. (a) Schematic structure of 1$T$- TiSe$_2$ in the top and side views. (b) STM image of fresh cleaved TiSe$_2$ surface (V$_s$ = 0.99 V, I = 0.27 nA). Insert: Atomically resolved STM image showing a Se atomic vacancy (V$_s$ = 0.1 V, I = 1.0 nA, scale bar = 0.5 nm). (c) STM image of monolayer vacancy islands on 1T-TiSe$_2$ after annealing (V$_s$ = 0.70 V, I = 0.31 nA). Insert: Atomically resolved STM image of fresh caved TiSe$_2$ surface (V$_s$ = 0.06 V, I = 1.29 nA, scale bar = 2 nm). (d) A zoom-in STM image of a single triangular vacancy monolayer island (V$_s$ = 0.99 V, I = 0.4 nA). Insert: Line profile crossing a triangular vacancy monolayer island marked by the solid blue line.

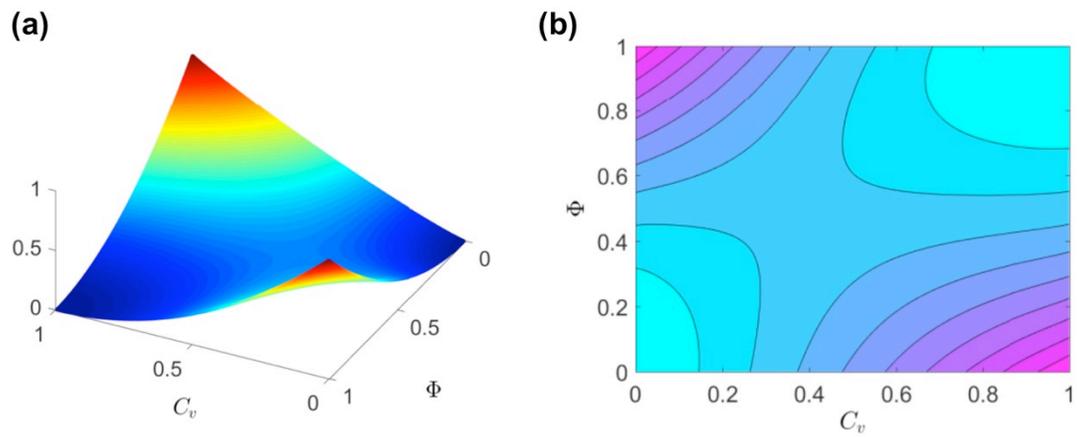

FIG. 2. (a) 3D representation of the bulk free energy of the system. (b) 2D contour plot of the bulk free energy.

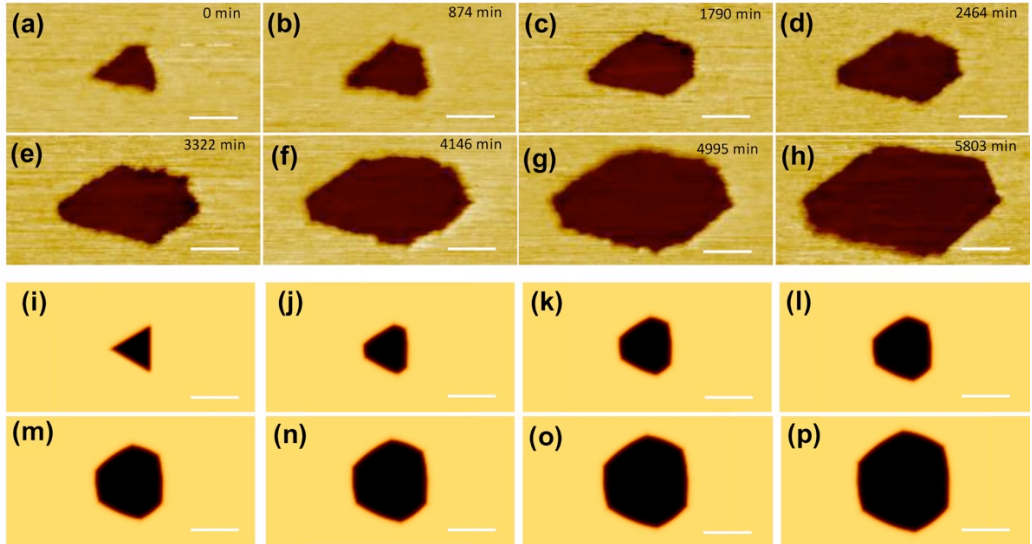

FIG. 3. STM images (a-h) and phase-field simulations (i-p) of the time evolution of a monolayer vacancy island on the surface of TiSe$_2$ under continuous scanning (Scale bar: 10 nm, $V_s$ = 1.0 V, I = 1.8 nA).

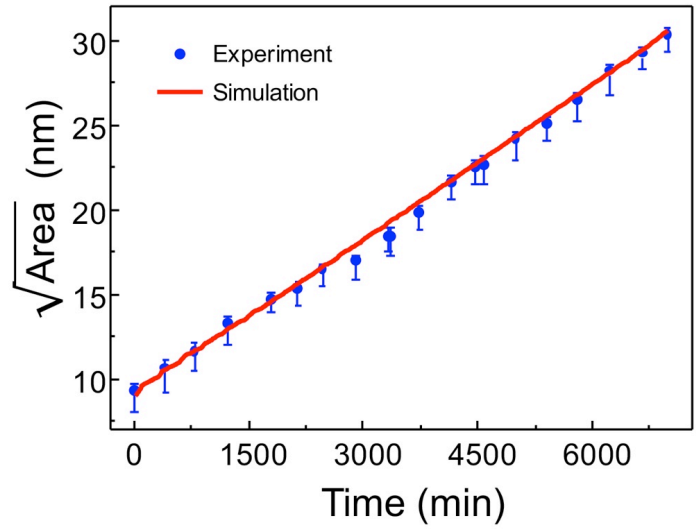

FIG. 4. Square root of area of a monolayer vacancy island vs. time ($V_s$ = 1.0 V, I = 1.8 nA). The solid line is result of a phase-field simulation with a preferential vacancy source term located at the vacancy island edges.

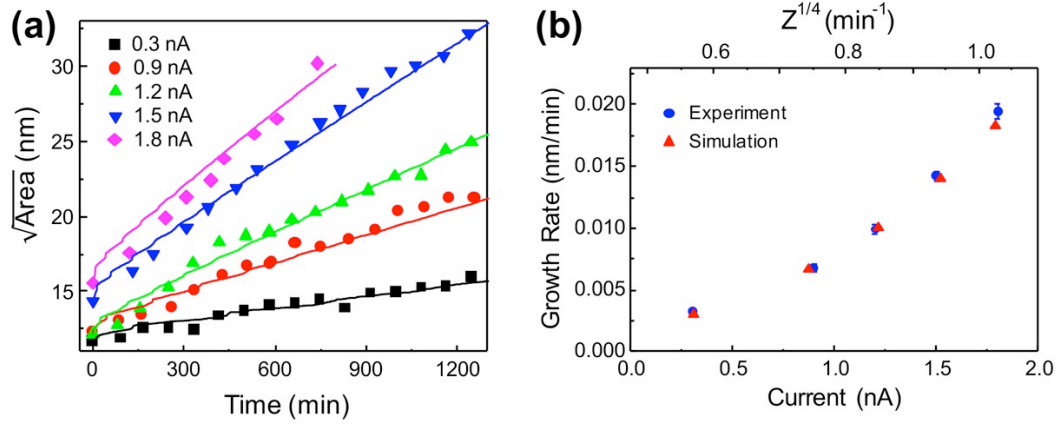

FIG. 5. (a) Time evolution of square root of area of monolayer vacancy islands at various tunneling currents ($V_s$ = 1.0 V), symbols represent experimental results and solid lines represent theoretical simulations. (b) Growth rate as a function of the tunneling current (blue circles) and plot comparing the growth rate against the ¼ power of the phase-field source constant $Z$ (red triangles).

**Acknowledgements**

C.T., H.Z. and C.C. acknowledge the financial support provided for this work by the U.S. Army Research Office under the grant W911NF-15-1-0414. K.H.B, N.O-O. and S.V. acknowledge financial support for NSERC of Canada, FQRNT of Quebec, and CONACYT of Mexico, as well as the computational resources provided by Compute Canada and Calcul-Quebec. N.O-O. also acknowledges the following financial assistance award 70NANB14H012 from U.S. Department of Commerce, National Institute of Standards and Technology as part of the Center for Hierarchical Materials Design (CHiMaD). L.J. acknowledges National Natural Science Foundation of China (No.51372134, No.21573125).